\def\be{\begin{equation}}
\def\ee{\end{equation}}
\def\bea{\begin{eqnarray}}
\def\eea{\end{eqnarray}}
\def\la{\langle}
\def\ra{\rangle}
\def\l{\langle\!\langle}
\def\r{\rangle\!\rangle}
\documentclass[prb,onecolumn,groupedaddress,preprint]{revtex4}
\usepackage{epsfig,graphicx}
\usepackage{bm}
\usepackage{dcolumn}
\usepackage{amssymb}

\begin{document}
\title{Conditional statistics of electron transport in interacting nanoscale conductors}

\author{Eugene V. Sukhorukov$^1$, Andrew N. Jordan$^2$, \\ Simon Gustavsson$^3$,  Renaud Leturcq$^3$, Thomas Ihn$^3$ \& Klaus Ensslin$^3$}

\affiliation{
${}^1$ D\'epartement de Physique Th\'eorique, Universit\'e de Gen\`eve,
        CH-1211 Gen\`eve 4, Switzerland\\
${}^2$ Department of Physics and Astronomy, University of Rochester, Rochester, New York 14627, USA\\
${}^3$ Solid State Physics Laboratory, ETH Zurich, CH-8093 Zurich, Switzerland}

\begin{abstract}
\end{abstract}

\maketitle
{\bf Interactions between nanoscale semiconductor structures form the basis for charge detectors in the solid state.
Recent experimental advances have demonstrated the on-chip detection of single electron transport \cite{Lu,Fuj,By,field,elzerman,schleser,vander,sprinzak}  through a quantum dot (QD).   The discreteness of charge in units of $e$ leads to intrinsic fluctuations in the electrical current, known as shot noise \cite{bb}.   To measure these single-electron fluctuations a nearby coherent conductor, called a quantum point contact (QPC), interacts with the QD and acts as a detector \cite{field,elzerman,schleser,vander,sprinzak}.  An important property of the QPC charge detector is noninvasiveness: the system physically affects the detector, not visa-versa.   Here we predict that even for ideal noninvasive detectors such as the QPC, when a particular detector result is observed, the system suffers an informational backaction, radically altering the statistics of transport through the QD as compared to the unconditional shot noise.  We develop a theoretical model to make predictions about the joint current probability distributions and conditional transport statistics.  The experimental findings reported here demonstrate the reality of informational backaction in nanoscale systems as well as a variety of new effects, such as conditional noise enhancement, which are in essentially perfect agreement with our model calculations.  This type of switching telegraph process occurs abundantly in nature, indicating that these results are applicable to a wide variety of systems.}

Noise is generally due to randomness, which can be classical or quantum in nature.  Telegraph noise, where there is random switching between two stable states \cite{machlup}, originates from such diverse origins as thermal activation of an unstable impurity \cite{ss1,ss2,ss3},  nonequilibrium activation of a bistable system \cite{act1,act2,act3}, switching of magnetic domain orientation \cite{mag1,mag2,mag3}, or a reversible chemical reaction in a biological ion channel \cite{bio}.

In nanoscale conductors, where charge motion is quantum-coherent over distances comparable to the system size, shot noise and telegraph noise have recently been shown to be two sides of the same coin \cite{schleser,vander,gust,fuji}.   A QD is sufficiently small that it is effectively zero-dimensional, and behaves as an artificial atom, holding a small number of electrons.  Figure 1{\bf (a)} shows the sample used in the experiment reported here. The QD is marked by the dotted circle \cite{fuhrer}. An extra electron can tunnel into the QD from the source lead (S), stay in the QD for a random amount of time, and then tunnel out into the drain lead (D) if the applied voltage bias exceeds the temperature.  This single-electron transport produces a fluctuating electrical current.   In order to detect the statistical properties of this current, a sensitive electrometer with a bandwidth much higher than the 
tunneling rates is required.  The electrometer is a nearby QPC that is capacitively coupled to the QD via the Coulomb interaction.  The voltage biased QPC detector transports many electrons through a narrow constriction in the surrounding two-dimensional electron gas (represented with an arrow).  The resistance of the QPC is susceptible to changes in the surrounding electrostatic environment, and can therefore be used to sense the presence (or absence) of an extra electron on the QD \cite{field}.  When the extra electron tunnels into or out of the QD, the current $I$ flowing through the QPC switches between two different values [see Fig.~1{\bf (b)}].  Therefore, the shot noise of the QD current $J$ (randomness in the number of switches in a given time interval) is intimately linked with telegraph noise in the QPC current $I$ (randomness of duration time in each current value).

\begin{figure}
\centering
 \includegraphics[width=10cm]{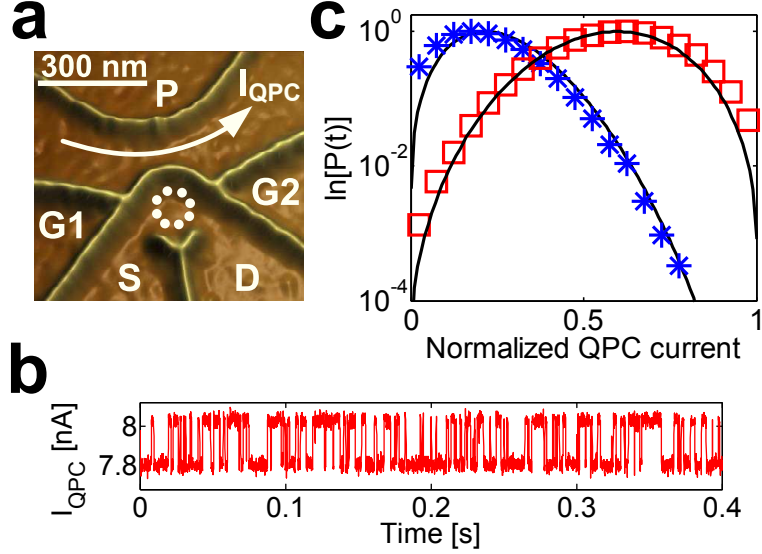}
 \caption{{\bf Nanodevice and current fluctuations.}  {\bf (a)} Quantum dot/quantum point contact structure fabricated using scanning probe lithography
\cite{fuhrer} on a $\mathrm{GaAs/Al_{0.3}Ga_{0.7}As}$ heterostructure containing a two-dimensional electron gas 34 $nm$ below the surface.  Ballistic electron flow through the QPC is indicated with an arrow, and the quantum dot is indicated with a dotted circle.  Individual electron current is induced by applying a voltage bias between source (S) and drain (D).   {\bf (b)} The flow of single electron current is deduced by the presence of switching in the QPC current, which acts as an on-chip electrometer. {\bf (c)} Probability distribution of charge transmitted through the QPC, plotted on a logarithmic scale. Data set A is represented by blue asterisks and data set B is represented by open red boxes.  The solid lines are theoretical prediction of a universal ellipse for both configurations, with no fitting parameter. Slight deviations from the ellipse at the ends of the interval are due to small-amplitude noise fluctuations in the individual current levels \cite{us}.}
\end{figure}

We now discuss the experimental procedure.   From a measurement such as shown in Fig. 1{\bf (b)}, one can directly determine the rates $\Gamma_{1,2}$ for electrons tunneling into and out of the QD \cite{schleser,naaman}. 
The tunneling rates are controlled by tuning the voltages applied to gates $G1$ and $G2$. The data presented here was taken at two
different gate voltage configurations.  Configuration A is characterized by $\Gamma_1=160~\mathrm{Hz}$, $\Gamma_2=586~\mathrm{Hz}$, and configuration B is characterized by $\Gamma_{1}=512~\mathrm{Hz}$, $\Gamma_{2}=345~\mathrm{Hz}$.  For each configuration, we collected traces of length $T=700~\mathrm{s}$, containing around $10^5$ tunneling events.  More details about the sample characteristics and the experimental methods can be found in the Supplementary Information.

In the following, we develop a theoretical model for this experiment. Ideal switching of the measured detector signal between two noiseless values $I_{1,2}$ are identified with the two current levels experienced by the QPC.  When an electron enters the QD, the current switches from $I_1$ to $I_2$, and when the electron leaves the QD, the reverse switch happens.  The number of ``down'' or ``up'' switches $M$ in a given time trace of duration $t$, is identified with the number of different transport electrons that occupy the QD in that time interval, naturally defining a QD current variable $J=M/t$ (we set $e=1$ to count in single electron charge units).  The analogous number of electrons $N$ passed by the QPC in this same time interval defines the QPC current variable $I=N/t$.   The assumption of noiseless current levels implies that $I_1 \le I(t) \le I_2$, while the unidirectional nature of the QD transport implies that $0 \le J(t) < \infty$  (see Supplementary Information for justification).   Stochastic, statistically independent quantum tunneling into and out of the QD is described with rates $\Gamma_{1,2}$ \cite{gust}.  For later convenience, we define the average and difference variables $I_0 = (I_1+I_2)/2, \Gamma_0 = (\Gamma_1 + \Gamma_2)/2, \Delta I =(I_2-I_1)/2, \Delta \Gamma = (\Gamma_2 - \Gamma_1)/2$.   This model is capable of describing a host of phenomena in many fields of science.

Taken alone, each side of the random process may be characterized mathematically with the probability distributions $P(I,t), P(J,t)$ of finding a given number of electrons transmitted in a given time, or equivalently, all current cumulants $\l I^n \r, \l J^m\r$ (see Supplementary Information for a discussion of these statistical quantities).  This catalog of cumulants gives a unique signature of the particular electronic conductor, and is also known as full counting statistics \cite{levitov}.  For example, the first two cumulants are the average current $\la I\ra$, and the shot noise power $\l I^2\r = \int dt \la \delta I(t) \delta I(0) \ra$, where $\delta I = I(t) -\la I \ra$.  For the QPC and QD, the average current $\la I\ra, \la J \ra$ and shot noise power $\l I^2\r, \l J^2\r$, are given respectively by
\bea
\la I \ra &=& (I_1 \Gamma_2 +I_2 \Gamma_1)/(2 \Gamma_0), \quad \l I^2\r =  (\Delta I)^2 \Gamma_1 \Gamma_2/\Gamma_0^3, \nonumber \\
\la J\ra &=& \Gamma_1 \Gamma_2/(2 \Gamma_0), \quad  \l J^2 \r=\la J \ra (\Gamma_1^2 +\Gamma_2^2)/(4 \Gamma_0^2).
\label{firsttwo}
\eea
In the limit of small switching rates $\Gamma_{1,2} \rightarrow 0$ the current and noise of the QD vanish, while the noise of the QPC
actually diverges.

However, a simple specification of the counting statistics of each conductor individually misses the important fact that the two conductors are strongly correlated by the Coulomb interaction between them.
The simplest measure of the correlation between the systems is the cross-correlation $\l I J \r  = \int dt \la \delta I(t) \delta J(0) \ra$, given by
\be
\l I J \r =(\Delta  I \Delta \Gamma)  \Gamma_1 \Gamma_2/\Gamma_0^2.
\label{cc}
\ee
This correlator is approximately constant under scaling of the switching rates, compromising between the behavior of either noise individually (\ref{firsttwo}). Result (\ref{cc}) implies that the two currents may be either positively or negatively correlated, depending on the
sign of $\Delta \Gamma \Delta I$.  This effect has a simple physical interpretation:  Taking $\Delta I>0$, if $\Gamma_2 > \Gamma_1$, then the system
typically spends more time in state 1 than in state 2.  The current $J$ is increased by adding another ``up'' and ``down'' switch to the current trace.
This new event typically divides a long time interval spent in $I_1$ into two segments, subtracting a short interval from $I_1$ and
adding it to $I_2$, thereby increasing the current $I$. The same argument applied with $\Gamma_2 < \Gamma_1$ leads to a decrease in $I$ given an increase in $J$, explaining the sign of the cross-correlation function.

Going beyond the first two cumulants, the full QPC current distribution was predicted to have an elliptical shape\cite{us} as a function of the current
\be
\log P(I)/t=-({\cal G}_1 - {\cal G}_2)^2/(2 \Delta I),
\label{ellipse}
\ee
where ${\cal G}_{1,2} = \sqrt{\Gamma_{1,2} \vert I - I_{2,1} \vert}$. This prediction is experimentally confirmed in Fig.~1{\bf (c)} for data sets A (blue asterisks) and B (red boxes).  We stress that the solid ellipses are taken directly from Eq.~\ref{ellipse} with no fitting. The generating function for the QD statistics was also found \cite{BN}, and the first few QD current cumulants were experimentally verified \cite{gust}.

In order to specify the statistical correlation between the conductors, we introduce the joint counting statistics of both conductors.
More specifically, the correlations may be quantified by the joint probability distribution $P(I,J,t)$ of finding current $I$ and current $J$ in a time $t$ (equivalently, all cross cumulants $\l I^n J^m \r$), or may also be specified by the conditional distribution functions $P(I\vert J)$ or $P(J \vert I)$, the probability of observing one current, given an observation of the other. These distributions are all related to one another by
$P(I,J) = P(I\vert J) P(J)= P(J\vert I) P(I)$, where the last equality is an expression of Bayes' theorem.  The conditional distributions (and their associated conditional cumulants) play a key role in the informational approach to detection \cite{korotkov}.

\begin{figure}
\centering
 \includegraphics[width=10cm]{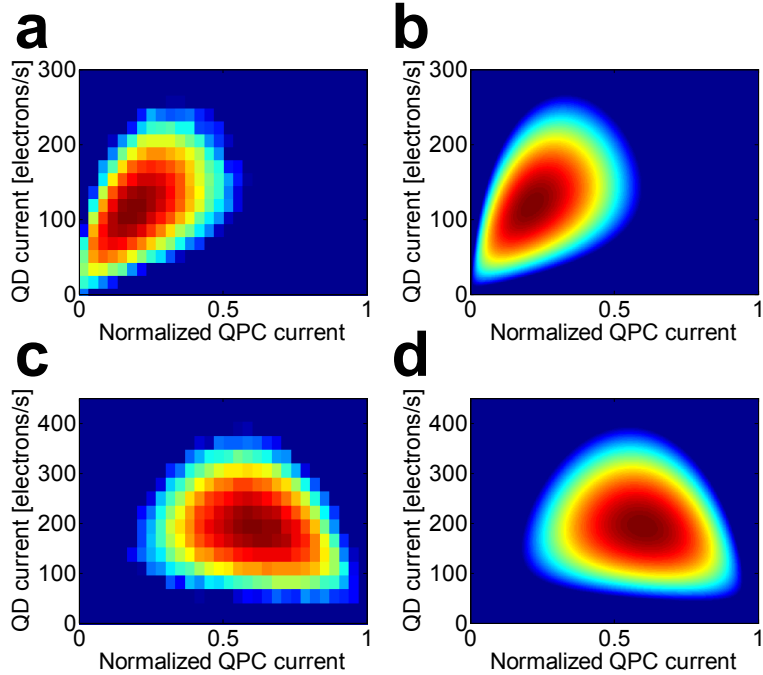}
 \caption{{\bf Joint probability distributions.}  The logarithm of the joint probability distribution of detecting QPC current (x-axis) and QD current (y-axis) is given as a color density plot, where red indicates high probability, and blue indicates low probability.  {\bf (a), (c)} Experimental construction for data set A and B respectively. The experimental probability distributions were generated by splitting the data into a large number of subtraces, each containing on average seven tunneling events. {\bf (b), (d)} Theoretical prediction for data set A and B respectively, see Supplementary Information for an explicit formula.}
\end{figure}
From the model described above, the complete characterization of the system/detector fluctuation statistics may be obtained from conditional master equation formalism.    The statistical cumulants of the current fluctuations are given with the help of a generating function $H(\lambda, \chi)$, such that the cross-cumulants are given simply by taking derivatives, $\l I^n J^m \r = \partial^n_\lambda \partial^m_\chi H(0,0)$.   From the mathematical derivation given in the Supplementary Information, we find that the joint generating function is given by
\be
H(\lambda, \chi) = \lambda I_0 - \Gamma_0 + \sqrt{(\lambda \Delta I - \Delta \Gamma)^2 + \Gamma_1 \Gamma_2 \exp \chi},
\label{gen}
\ee
generalizing previous results \cite{BN,us}.   The function $H(0, \chi)$ generates the current cumulants of the QD \cite{BN}, while $H(\lambda, 0)$ generates the current cumulants of the QPC \cite{us}.  Results (\ref{firsttwo}) and (\ref{cc}) follow from (\ref{gen}).   The joint generating function (\ref{gen}) is directly related to the joint probability distribution of measuring current $I$ and current $J$ [see Supplementary Eq.~(6)].  The logarithm of this distribution has been measured and given in Fig.~2{\bf (a), (c)} for configurations A and B respectively.  The theoretical prediction for this quantity is given in Fig.~2{\bf (b), (d)} with striking agreement.

Having described the joint statistical properties of both currents, we now return to the detection question.
It is important to distinguish between physical backaction from statistical/informational backaction.
The noninvasive QPC detector changes its physical current state depending on whether or not the QD is occupied by an extra electron, while the physical dynamics of the QD is unaffected by the state of QPC.  However, by observing a particular outcome of the detector variable ($I$), this leads to conditional (Bayesian) backaction of the detector on the system variable ($J$).  This informational backaction, or constrained randomness, manifests itself in a variety of novel effects. We introduce the concept of {\it conditional counting statistics}:  The statistical current fluctuations of one system, given the observation of a given current in the other.  These statistics may be calculated from the joint
generating function (\ref{gen}) (see Supplementary Information), giving a mixed generator ${\cal H}_1(I, \chi)$ of the (normalized) conditional statistics of the QD, given the observation of a current $I$,
\be
{\cal H}_1 = 
\Omega (e^{\chi/2} -1),\quad \Omega  = \frac{\sqrt{\Gamma_1 \Gamma_2}}{\Delta I} \sqrt{(I-I_1)(I_2 - I)}.
\label{mixed1text}
\ee
Taking derivatives, all conditional cumulants are given by 
$\l J^n \r_c = \Omega/2^n$, 
yielding a set of universal semi-circles as a function of the current $I$.  At the endpoints of the interval, the QPC current is observed to remain in $I_1$ or $I_2$, and therefore there can be no QD current, or any associated noise.  The conditional QD current cumulants all have a maximum at $I_0$.  The conditional current maximum $\sqrt{\Gamma_1 \Gamma_2}/2$ is always larger than the unconditional current $\la J\ra$.
Fig.~3 {\bf (a), (b)} compare the experimental values of the first two conditional cumulants to the theoretical semicircles, with excellent agreement.
The distribution described by the conditional cumulants (\ref{mixed1text}) is a Poissonian distribution with a generalized rate $\Omega$, and effective charge $e^{\ast}=e/2$, showing a radical change when compared with the unconditional distribution \cite{BN}.

Turning our perspective around, we can pose similar questions about the conditional detector statistics, given an observation of the system current $J$.
While the generating function ${\cal H}_2(\lambda, J)$ for these statistics may be found analytically (see Supplementary Information), we focus on the conditional current $\la I \ra_c$, and the conditional noise $\l I^2\r_c$, given by
\be
\la I \ra_c =  I_0 - \frac{\Delta I \Delta \Gamma}{J + S}, \quad
\l I^2 \r_c =  \frac{(\Delta I)^2 \, J}{S(J + S)},
\label{condI}
\ee
where $S = \sqrt{J^2 + (\Delta \Gamma)^2}$.  These conditional cumulants are experimentally calculated and compared with Eq.~(\ref{condI}) in Fig. 3{\bf (c),(d)}.   As a function of the QD current $J$, the conditional current tends to either $I_1$ or $I_2$ as $J\rightarrow 0$, depending on the sign of $\Delta \Gamma$.  This corresponds to the most likely detector current configuration in the event of no switches observed:  the QPC current stays in one value, also implying that the system becomes noiseless in this limit.  This is easily seen in (\ref{condI}) because $\l I^2 \r_c$ is proportional to $J$.   The exception to this rule is the perfectly symmetric situation $\Gamma_1 = \Gamma_2$, where the QPC conditional average current is $I_0$.  This situation corresponds to rare symmetric switching between the states, whose effective rate is the conditional QD current $J$.  The corresponding QPC conditional noise actually diverges in this limit, because the effective switching rate is vanishing. This effect, where the noise in one system (monitored by another) can be dramatically larger than the unmonitored noise, we refer to as {\it conditional noise enhancement}.  The same effect persists in the asymmetric situation, and the maximum of the conditional noise occurs at $J^2 = (\Delta \Gamma)^2 (\sqrt{5}-1)/2$.  
In order that the conditional noise peak exceed the unconditional noise, the ratio $R=|\Delta \Gamma| \Gamma_1 \Gamma_2/\Gamma_0^3$ must be less than $[(\sqrt{5}-1)/2]^{5/2} \approx 0.3$.   For data set A and B, $R_A\approx 0.38$ and $R_B \approx 0.19$, so only data set B exhibits conditional noise enhancement.   In the opposite limit, $J\rightarrow \infty$, the conditional current tends to $I_0$, and the noise tends to zero.  This situation corresponds to rapid symmetric switching between the current states, whose effective rate is again controlled by $J$.  In both limits, the typical dynamics of the telegraph process gets completely taken over by the transport condition.


\begin{figure}
\centering
 \includegraphics[width=10cm]{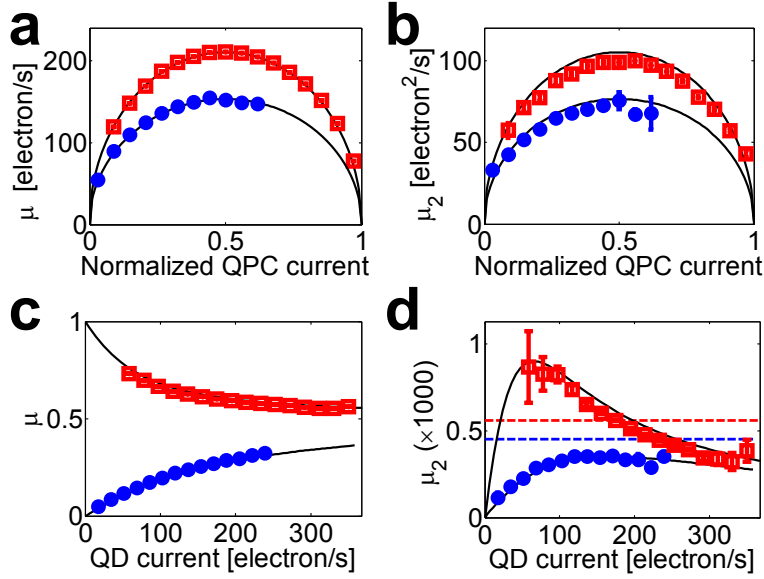}
 \caption{{\bf Conditional current cumulants.}  {\bf (a)} Conditional QD current $\la J\ra_c$ plotted as a function of $I$.
 {\bf (b)} Conditional QD noise $\l J^2\r_c$ plotted as a function of $I$.  Solid line is theory from Eq.(\ref{mixed1text})
 describing universal semi-circles.  Solid blue dots and open red squares denote data set A and B respectively. {\bf (c)} Conditional QPC current $\la I\ra_c$ plotted as a function of $J$.  {\bf (d)} Conditional QPC current $\l I^2\r_c$ plotted as a function of $J$.  Solid line in {\bf (c)} and {\bf (d)} is theory, given by Eq.~\ref{condI}.  In {\bf (d)} the horizontal dashed lines indicate the unconditional noise level, illustrating the effect of ``conditional noise enhancement'' for data set B described in the text.  The experimental statistics were generated by splitting the data into a large number of subtraces, each containing on average four tunneling events.
 }
\end{figure}


\begin{acknowledgments}
We acknowledge the support of the Swiss National Science Foundation. We thank
M. Reinwald and W. Wegscheider at the Institut f\"ur experimentelle und angewandte Physik Universit\"at Regensburg, Germany
for fabricating the wafers used in this experiment.
\end{acknowledgments}

\bibliography{conditional_condmat}  

\pagebreak

\noindent {\bf Supplementary Information} \\

\noindent {\bf ``Conditional statistics of electron transport in interacting conductors''} \\

\noindent E. V. Sukhorukov, A. N. Jordan, S. Gustavsson, R. Leturcq, T. Ihn \& K. Ensslin\\

\noindent {\bf Sample Fabrication}

Figure 1{\bf (a)} shows the sample used in the experiments reported here. The structure was fabricated using scanning scanning probe lithography on a $\mathrm{GaAs/Al_{0.3}Ga_{0.7}As}$ heterostructure containing a two-dimensional electron gas 34 nm below the surface. 
\\

\noindent {\bf Sample Characteristics}

The measurements were performed in a $^3$He/$^4$He dilution refrigerator with an electron temperature of about 190 mK, as determined from the width of thermally broadened Coulomb blockade resonances. The charging energy of the QD is 2.1 meV and the mean level spacing is $200$ $-$ $300~\mu eV$.
\\

\noindent {\bf Experimental Method}

The conductance of the QPC, $G_{QPC}$, was tuned close to  $0.5 \times 2 e^2/h$. We apply a dc bias voltage between the source and the drain
of the QPC, $V_{QPC} = 250~\mu V$, and measure the current through the QPC, $I_{QPC}$, which depends on the number of electrons $N$ in the QD.  The current signal was digitized with a sampling frequency of $100~\mathrm{kHz}$, thereafter software filtered at $4~\mathrm{kHz}$ using a 8th order Butterworth filter.  To avoid tunneling due to thermal fluctuations, the QD was voltage biased with a bias much larger than the temperature.
\\

\noindent {\bf Supplementary Discussion:  Introduction to the basic concepts of transport statistics}\\

Central to the discussion of transport statistics is the concept of the probability distribution $P(Q,\Delta t)$ of transporting $Q$
charges in a time $\Delta t$.   Defining the {\it moment generating function} $G$ of the distribution
\be
G(\lambda, \Delta t) = \sum_Q e^{\lambda Q} P(Q, \Delta t), \nonumber
\ee
the $k$th moment $\la Q^k\ra(\Delta t)$ of the distribution $P(Q, \Delta t)$ is given by simply taking derivatives with
respect to $\lambda$, the generating variable,
\be
\la Q^k\ra(\Delta t) = \partial_\lambda^k G(0) = \sum_Q Q^k P(Q, \Delta t). \nonumber
\ee
If $k=0$, then the result is 1 because the distribution is normalized.
We further define the {\it cumulant generating function}
as $S(\lambda, \Delta t) = \log G(\lambda, \Delta t)$, so the {\it cumulants} of the distribution $\l Q^k \r$ are defined by
\be
\l Q^k \r(\Delta t) = \partial_\lambda^k S(0, \Delta t). \nonumber
\label{chargecum}
\ee
For example, $\l Q \r = \la Q \ra, \l Q^2 \r = \la (Q -\la Q\ra)^2 \ra, \l Q^3 \r = \la (Q -\la Q\ra)^3 \ra, 
\l Q^4 \r = \la (Q -\la Q\ra)^4 \ra - 3 \la (Q -\la Q\ra)^2 \ra$, etc.
The distribution may be expressed in terms of the cumulant generating function as
\be
P(Q, \Delta t) = \int \frac{d\lambda}{2\pi} \exp[-\lambda Q + S(\lambda, \Delta t)] \nonumber
\label{fourier}
\ee
where the integration is over $[0, 2\pi i]$.

For a random transport process to be {\it Markovian} means that the time evolution at time $t+ \Delta t$ is only
determined by the state of the system at time $t$, where $\Delta t$ is larger than any microscopic correlation time scale.
Markovian noise is {\it stationary} if the probability of transmitting charge $Q_1$ in time $\Delta t_1$ followed by charge $Q_2$ in time $\Delta t_2$ through a conductor is simply given by the product of the probability distributions, so the two events are statistically independent in time.
This statistical independence implies that the cumulant generating function for transporting $Q$ charges in time $\Delta t = \Delta t_1 +\Delta t_2$
obeys the relation
$S(\lambda, \Delta t) = S(\lambda, \Delta t_1) + S(\lambda, \Delta t_2)$.  This equation
implies that the charge cumulant generator $S(\lambda, \Delta t)$ must be linear in $\Delta t$ if the Markovian random process is stationary,
\be
S(\lambda, \Delta t) = \Delta t\, H(\lambda). \nonumber
\label{H}
\ee
We define $H(\lambda)$ as the generating function of the current cumulants.
The reason cumulants of current play a central role in the statistical characterization of transport, is the fact that the
cumulants of charge must also be linear in time.  Therefore,
the cumulants of current $\l I^k \r$ are a {\em time-independent} characterization of the statistics of the
transport, given by
\be
\l I^k \r = \l Q^k \r /\Delta t = \partial_\lambda^k H(0).  \nonumber
\ee
\\

\noindent {\bf Supplementary discussion:  Details of the Mathematical Derivations} \\

In order to describe the transport statistics of the interacting QD/QPC system, we begin
with the simplest case of no transport through the QD.
The statistical properties of the QPC are described by the current cumulant generating
functions $H_\alpha(\lambda) = \sum_k \lambda^k \l I^k_\alpha\r/k! $, where $\alpha=1,2$ indicates
whether the QD is occupied or empty. These functions describe the intrinsic fluctuations of the QPC in the absence of any switching.

To access the statistics of the QD and QPC when switching with rates $\Gamma_{1,2}$ is allowed,
we now derive the joint generating function of both conductors, based on conditional master equation formalism.
From the model described in the main text, transport through the quantum dot occurs through
electrons tunneling from the left reservoir, onto the QD, and off through the right reservoir.
If the number of electrons in the right (left) reservoir on either side of
the QD is taken to be $N_R (N_L)$, then whether the current $I$ is either $I_1$ or $I_2$ is controlled by the
resident charge of the QD, while the current through the QD is controlled by the transfered charge.  The
transfered charge $M$ is given by counting the change in the symmetric combination $(N_R - N_L)/2$ from $t=0$,
while the resident charge is given by the change in $ -(N_R+N_L)$ from $t=0$ (we note that it is important to
choose these two independent variables to avoid introducing superfluous correlations into the system).  We introduce the joint probability $P_\alpha(N,M,t)$ of transmitting $N$ electrons through the QPC, and $M$ electrons through the QD in the time $t$ in state $\alpha$.  The mixed generator ${\cal U}_\alpha(\lambda, M, t) = \sum_N e^{\lambda N} P_\alpha(N,M,t)$ satisfies the master equation
\bea
\partial_t {\cal U}_1(M) &=& (H_1 -\Gamma_1) {\cal U}_1(M) + \Gamma_2 {\cal U}_2(M+1/2), \label{cmea}\\
\partial_t {\cal U}_2(M) &=& (H_2 -\Gamma_2) {\cal U}_2(M) + \Gamma_1 {\cal U}_1(M+1/2).
\label{cmeb}
\eea
This equation describes the time rate of change of the probability of occupying state $\alpha$ in terms of incoming and outgoing
contributions.  The current transfered through QPC state 1 is counted by the first term in (\ref{cmea}), while the possibility
of switching to state 2 is accounted for in the second term (\ref{cmea}) where $N_L$ decreases by one, accompanied by a change in $M$.
Analogously, the current transfered through QPC state 2 is counted by the first term in (\ref{cmeb}), while the possibility
of switching to state 1 is accounted for in the second term (\ref{cmeb}) where $N_R$ increases by 1, accompanied by a change in $M$.

Introducing the full generator $U_\alpha(\lambda, \chi, t) = \sum_M e^{\chi M} {\cal U}_\alpha(\lambda, M, t)$, the full
conditional master equation is given by the matrix equation $\partial_t U = {\bf H} U$, where
\be
{\bf H}(\lambda, \chi) = \left(\begin{array}{cc}
 H_1(\lambda) - \Gamma_1  & \Gamma_2 e^{\chi/2}\\
\Gamma_1 e^{\chi/2} & H_2(\lambda) - \Gamma_2 \end{array}\right).
\label{condme3}
\ee
For times larger than $\Gamma_{1,2}^{-1}$, the systems relaxes to the stationary distribution,
controlled by the largest eigenvalue $H(\lambda, \chi)$, so that $\sum_\alpha U_\alpha \sim \exp[t H(\lambda, \chi)]$.
Considering just the noise power (second cumulant) of the QPC alone, $\partial_\lambda^2 H(0,0)$, the result is 
\be
\l I^2\r = \sum_{\alpha=1,2} \l I^2 \r_\alpha P_\alpha +  (\Delta I)^2\Gamma_1\Gamma_2/\Gamma_0^3,
\label{comp}
\ee
where $P_{1,2} =  \Gamma_{2,1}/(2\Gamma_0)$ are the probabilities of occupying state $\alpha$, and $\l I^2 \r_\alpha = H_\alpha''(0)$ is the
microscopic (shot) noise of the QPC in state $\alpha$.  The comparatively long switching times implies that the telegraph contribution to (\ref{comp})
dominates over the QPC shot noise, and therefore we may neglect the QPC shot noise.  This amounts to replacing $H_\alpha(\lambda) \rightarrow \lambda I_\alpha$, which recovers Eq.~(4) of the main text.

From the definition of the generating function, the probability distribution of transmitted charge may
be extracted by Fourier transforming on both variables,
\be
P(N,M, t) = \int \frac{d\chi d \lambda}{(2\pi)^2} \exp[ t H(\lambda, \chi) - \chi M - \lambda N],
\label{pnm}
\ee
where the integration is over the interval $[ 0, 2\pi i]$.
Replacing $N = I t, M=J t$ gives the distributions of the two currents.  In order to consider the
stationary limit, it is necessary to have $t > \Gamma_{1,2}^{-1}$, and we may therefore evaluate
the integral (\ref{pnm}) in the stationary phase approximation.
The dominant contribution to the joint probability distribution is then given by
\be
\log P(I,J) =t\, {\rm min}_{\lambda, \chi} [H(\lambda, \chi) - I \lambda - J \chi],
\ee
which takes the form of a Legendre transform, yielding the result
\be
\log P /t = -\frac{\Delta \Gamma}{\Delta I} (I- I_0) - { \Gamma_0} - 2 J \left[\log(2 J/ \Omega)-1\right], \quad
\Omega  = \frac{\sqrt{\Gamma_1 \Gamma_2}}{\Delta I} \sqrt{(I-I_1)(I_2 - I)}.
\label{dist}
\ee
The term inside the logarithm is responsible for the correlation between the two currents.
The mixed generating functions may be calculated by only Fourier transforming on one of
the above variables.  Integrating over the $\lambda$ variable only gives the first mixed form,
\be
{\cal H}_1(I, \chi) = -\frac{({\cal G}_1 - {\cal G}_2)^2}{2 \Delta I} + \Omega (e^{\chi/2} -1),
\label{mixed1app}
\ee
where ${\cal G}_{1,2} = \sqrt{\Gamma_{1,2} \vert I - I_{2,1} \vert }$.  The result in the main text Eq.~(5)
is given by normalizing the conditional distribution by subtracting off the $\chi=0$ contribution.
Equation (\ref{mixed1app}) recovers the result Eq.~(3) in the main text for the special case of $\chi=0$.  
The other conditional generating function is found by transforming on $\chi$ only, yielding the second mixed form
\be
{\cal H}_2(\lambda, J) = \lambda I_0 - \Gamma_0  + {\cal S} - J \log [2 J {\cal S}/\Gamma_1 \Gamma_2], 
\ee
where ${\cal S} =  J + \sqrt{J^2 + (\lambda \Delta I - \Delta \Gamma)^2} $.
The conditional cumulants (6) of the main text are found by taking derivatives of this generating function with respect to $\lambda$.

Applying the remaining Fourier transformation to either of the above forms yields the result (\ref{dist}), completing the circle.
We note that the conditional distribution functions may be obtained by Fourier transforming with respect to either
${\cal H}_1(I, \chi) - {\cal H}_1(I, 0)$ to obtain $P(J\vert I)$, or with ${\cal H}_2(\lambda, J) - {\cal H}_2(0, J)$ to obtain $P(I\vert J)$,
as a consequence of Bayes' theorem.

\end{document}